\documentclass[aps,prb,amsmath,amssymb,superscriptaddress,showpacs,twocolumn,floatfix]{revtex4}
\usepackage{amsfonts, relsize}
\usepackage{graphics}
\usepackage{graphicx}
\usepackage{hyperref}

\begin{document}

\title{Topological insulators in strained graphene at weak interaction}
\author{Bitan Roy}
\affiliation{ National High Magnetic Field Laboratory and Department of Physics, Florida State University, Florida 32306, USA}

\author{Igor F. Herbut}

\affiliation{ Max-Planck-Institut f\"ur Physik Komplexer Systeme, N\"othnitzer Stra$\beta$e 38, 01187 Dresden, Germany}
\affiliation{ Department of Physics, Simon Fraser University, Burnaby, British Columbia, Canada V5A 1S6}

\date{\today}

\begin{abstract}
The nature of the electronic ground states in strained undoped graphene at weak interaction between electrons is discussed. After providing a lattice realization of the strain-induced axial magnetic field we numerically find the self-consistent solution for the time reversal symmetry breaking quantum anomalous Hall order parameter, at weak second-nearest-neighbor repulsion between spinless fermions. The anomalous Hall state is obtained in both uniform and nonuniform axial magnetic fields, with the spatial profile of the order parameter resembling that of the axial field itself. When the electron spin is included, the time reversal symmetric anomalous spin Hall state becomes slightly preferred energetically at half filling, but the additional anomalous Hall component should develop at a finite doping.
\end{abstract}

\pacs{71.10.Pm, 71.70.Di, 73.22.Pr}

\maketitle

\vspace{10pt}

\section{Introduction}

Graphene, being a flexible membrane, is always wrinkled to a certain degree. One can think of the effect of the wrinkles on the quasirelativistic Dirac low-energy excitations in graphene as the random, static component of the non-Abelian (or ``axial") gauge field, which arises from modified hopping amplitudes in the tight-binding model, and which preserves the time reversal invariance. Wrinkles are typically randomly distributed, yielding roughly zero total flux of the axial field. If one deliberately bulges graphene, however, the total axial magnetic flux may become finite. In a recent experiment \cite{levy} graphene was deposited over a metallic substrate and cooled down. The mismatch of graphene's and substrate's compressibilities subjects the graphene layer then to a strain, resulting in a surprisingly  uniform axial magnetic field as high as $\sim 350$ T. We will argue here that these may be the ideal conditions for a possible observation of the dynamically induced quantum anomalous Hall (AH) and the quantum anomalous spin Hall (SH) states, the former predicted before in Ref.~\onlinecite{herbut-pseudocatalysis}. In so doing we will go beyond the usual Dirac approximation and present numerical evidence for the effect, even in the more general case of a spatially nonuniform axial field.

The gist of the effect lies in part in the index theorem, \cite{aharonov} which equally well applies to the axial and real magnetic fields: A finite arbitrary  field, in the continuum,  produces single-particle states at zero energy, proportional in number to the total flux. This zero-energy degenerate band in graphene is half filled, and it is energetically advantageous to split it and populate only the lower half. While the density of states does not distinguish between the real and axial magnetic fields, the spinor structure of the zero modes does: The zero modes at two inequivalent Dirac points (valleys) live on the same sublattice in the presence of axial magnetic fields, whereas they live on the complementary lattices in the real magnetic field. As a result, the following theorem follows, for spinless particles: Although real magnetic field allows the formation of various chiral symmetry breaking orders, in the presence of axial fields {\it only} the time reversal symmetry (TRS) breaking AH insulator can split the zero-energy subband.\cite{herbut-pseudocatalysis} In this paper we focus on the interaction effects in the axial field. The reader interested in the behavior in real field may consult an excellent recent review of the subject. \cite{goerbig}
\begin{figure}[htb]
\includegraphics[width=8.5cm, height=5.5cm]{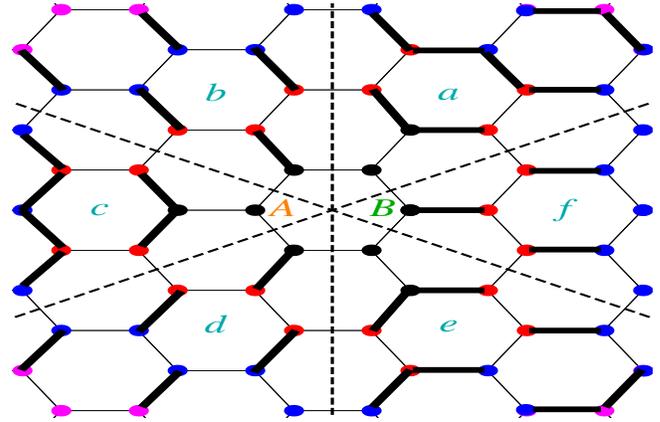}
\caption{(Color online) $\chi({\cal R})$ attached to each site; its value increases in the order black, red, blue, magenta. Thick bonds represent modified hopping amplitude. Sections $(a,c,e)$, and $(b,d,f)$  are connected by $2\pi/3$ rotations.}\label{phase}
\end{figure}

Since, at least without the spin, the only available instability to the system is the AH state, we use the simplest interacting Hamiltonian which yields the effect, with the repulsion only between the second-nearest neighbors. While this is admittedly only one component of the otherwise long-range Coulomb interaction, the others are, at least at the Hartree-Fock level, inert in the presence of the axial field. Chiral (``valley-rotational") symmetry, the breaking of which is usually the dominant instability in zero or finite true magnetic field,\cite{herbut-interaction, grushin} is here already broken by the axial field, which leaves the time reversal as the only remaining symmetry to be spontaneously broken by the interactions. It is then demonstrated that the instability towards the AH insulator happens in the presence of even a weak such repulsion, which, when strong, is thought to favor the formation of the AH state, even without any axial flux. \cite{raghu} Remarkably, the TRS breaking AH order parameter (OP) \cite{haldane} is found in our solution to be spatially distributed similarly to the axial flux itself, and would thus be close to uniform in  a uniform field in the experiment. We first present the detailed numerical results of our self-consistent calculation for the spinless fermions. With the restoration of spin, other ordered states, such as the spin polarized ferromagnetic state \cite{ashwin} and the anomalous SH insulator \cite{abanin} also become possible. Neglecting possible effects of the  Hubbard on-site interaction we find that, although degenerate at the mean-field level, fluctuations in this case slightly favor the SH state in the finite axial field and at weak coupling.

The rest of the paper is organized as follows. In the next section, we provide a lattice realization of axial magnetic fields in graphene, and present the energy spectrum of a system of noninteracting fermions, subject to uniform and nonuniform axial magnetic fields. In Sec. III, we discuss the role of the electron-electron interaction in strained graphene within the mean-field approximation, and show that an AH insulator can be realized even for arbitrary weak next-nearest-neighbor repulsion between the spinless fermions. We also present the scaling  of the AH OP and show that such ordering occurs in the presence of a uniform or nonuniform axial field. In Sec. IV the spin degrees of freedom of fermions is restored, and we study the competition between the AH and the SH insulators. We summarize our results and discuss some related issues in Sec. V. We also provide `Supplementary Material with additional numerical results.
\begin{figure}[htb]
\includegraphics[width=7.0cm,height=5.0cm]{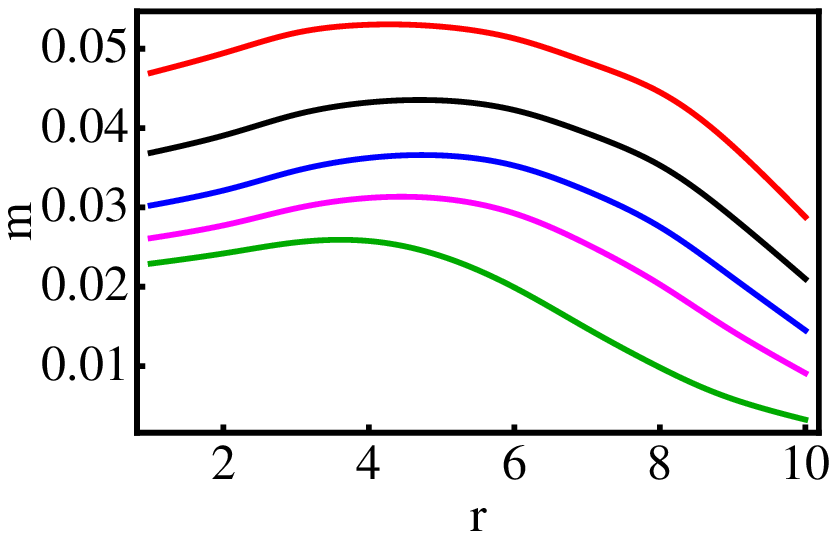}\vspace{0.3cm} \\
\includegraphics[width=7.0cm,height=5.0cm]{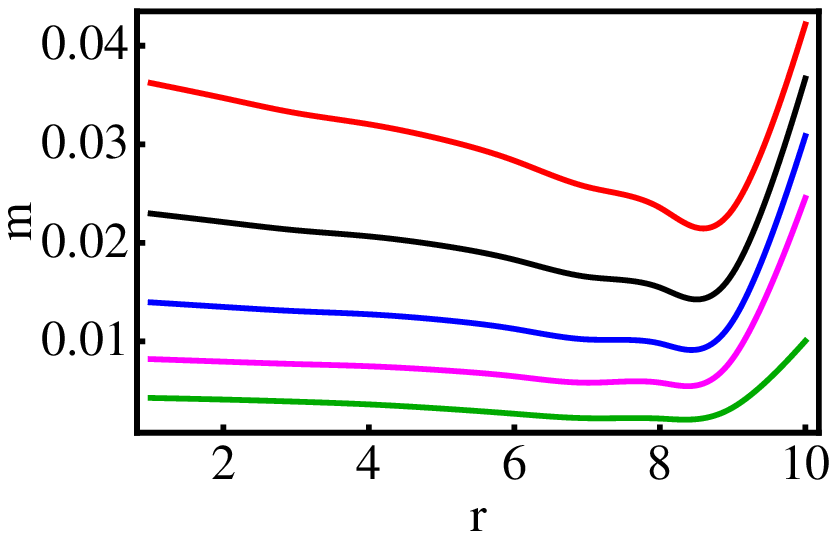}
\caption{(Color online) AH OP (in units of $t$) on A(top), B(bottom) sublattice in the presence of uniform axial field $b=0.025 b_0$. $b_0= \hbar/(e a^2)\approx 10^4$ T, is the field associated with graphene's lattice spacing $a \approx 2.5 \mathring{A}$. $\hbar$ is Plank's constant, $e$ is electronic charge. Interaction strength reads as $V=1.5,1.27,1.0,0.75,0.5$ from top to bottom. $r=n (1,2, \cdots)$ corresponds to $n$th ring around the center of the system\cite{lattice-geometry}.}\label{AHlattice}
\end{figure}

\section{Axial field on the lattice and free fermions}
To construct a lattice implementation of the axial field, we recall first the low-energy Dirac Hamiltonian in graphene. In the presence of an axial field, it may be written as \cite{herbut-pseudocatalysis, Jackiw}
\begin{equation}
H[a]= i \gamma_0 \gamma_i ( -i\partial_i- i \gamma_3 \gamma_5 a_i (\vec{x})) \equiv  e^{\chi(\vec{x})\gamma_0} H[0] e^{\chi(\vec{x})\gamma_0},
\label{continuumHamiltonian}
\end{equation}
 where $i=1,2$. Neglecting spin, the Dirac Hamiltonian acts on the four component fermion, defined as
\begin{equation}\label{spinor}
\Psi^\top(\vec{q}) =
( u (\vec{K} + \vec{q}), v (\vec{K}+ \vec{q}), u (-\vec{K}+ \vec{q}), v (-\vec{K}+ \vec{q}) ).
\end{equation}
$u$ $(v)$ is the annihilation operator on sublattice A (B), and $a_i (\vec{x}) = \epsilon_{ij} \partial_j \chi (\vec{x})$. The four-component Hermitian $\gamma$-matrices are $\gamma_0=\sigma_0\otimes\sigma_3,\gamma_1=\sigma_3\otimes\sigma_2,\gamma_2=\sigma_0\otimes\sigma_1, \gamma_3=\sigma_1\otimes\sigma_2, \gamma_5=\sigma_2\otimes \sigma_2$ \cite{herbut-interaction, kekule}. In the presence of an axial field, the emergent chiral $SU_c(2)$ symmetry of $H[0]$, generated by $\{\gamma_3,\gamma_5,i\gamma_3 \gamma_5 \}$, is reduced to $U_c(1)$, generated by  $i \gamma_3 \gamma_5$.\cite{pseudo-gauge-potential} The pseudo magnetic field is $b (\vec{x}) =\epsilon_{i j} \partial_i a_j(\vec{x})=\partial^2 \chi(\vec{x})$.

From Eq. (\ref{continuumHamiltonian}) one can write the zero energy states in the presence of axial field as
\begin{equation}
\Psi_{0,n}\left[a\right](\vec{x})\; \propto \; e^{-\chi(\vec{x})\gamma_0} \; \Psi_{0,n} \; \left[0\right](\vec{x}).
\end{equation}
The matrix $\gamma_0$ in the exponent alternates in sign on A and B sublattices. Since $\chi(\vec{x})$ is an increasing function of the distance, \cite{aharonov} only the sublattice A supports normalizable zero-energy states. Zero-energy states  on the sublattice B will diverge at infinity, i. e., in a finite system would be localized  near the boundary. This suggests an introduction of the axial gauge potential on lattice as the following modification of the nearest-neighbor hopping integrals:
\begin{equation}
t_{\alpha \beta} \;=\; e^{\chi(\alpha)} \: t \: e^{-\chi(\beta)},
\label{modhopping}
\end{equation}
where $\alpha \in A$, $\beta \in B$, and $t (=1)$ is the uniform hopping.\cite{huse}

Upon defining a quantity ${\cal R}$, counting the minimal number of bonds required to reach a particular site from the central hexagon, we assign $\chi({\cal R})$ to each site, depending on whether it belongs to the sublattice A or B,  in the following way: when ${\cal R}$ is odd, $\chi(A)>\chi(B)$, and when it is even, $\chi(A) = \chi(B)$. This is presented in Fig.~\ref{phase}. ${\cal R}$ here plays the role of the radial coordinate, and for all six sites in the central hexagon in Fig.~\ref{phase} ${\cal R}=0$, for example. Then along each bond with modified hopping amplitude $\chi(A)>\chi(B)$, in agreement with Eq.~(\ref{modhopping}). Such modification leaves the honeycomb lattice invariant under the $C_3$ symmetry, and an axial vector potential $\vec{a}=\left( a_{{\cal R}},a_\phi\right)\approx \left( 0,\partial\chi({\cal R})/\partial {\cal R} \right)$ is introduced in the system. If $\chi({\cal R})\propto {\cal R}^2$, the axial flux enclosed by the system is roughly proportional to its area, corresponding to an approximately \emph{uniform} axial magnetic field. On the other hand, a \emph{bell-shaped} axial field, localized around the center of the system, can be obtained by choosing $\chi({\cal R}) \propto \ln({\cal R})$.

Let us first take a spinless fermion hopping on a honeycomb lattice of 2400 sites,\cite{lattice-geometry} subject to uniform and nonuniform axial fields. Since the non-interacting Hamiltonian is bipartite, the diagonalization always gives a particle-hole symmetric spectrum, with a finite number of states localized close to the zero energy. With the bond configuration as in the above, we indeed find that the near-zero modes that are on the sublattice A are localized in the bulk of the system, while those that are on the sublattice B are localized near the system's boundary. Furthermore, when the field is roughly uniform, isolated windows of energy $W \sim t/10$ where the number of states increases with the axial flux appear symmetrically around zero. These states appear to form the first Landau level (LL), and we define the mean energy of these states, which scales roughly as $\sqrt{b}$, as the first LL energy ($E_1$). The lack of perfectly sharp LL quantization in our calculation is due to the finite size effects, and also to the local modification of the Fermi velocity\cite{bitan-hou-yang}.

\section{Interacting spinless fermions in axial magnetic field}

Next, we turn on the second-nearest-neighbor repulsive interaction ($V>0$) between the fermions, still kept spinless. This interaction is singled out because, as it will become clear shortly, it directly favors the only possible instability in the (spinless) system, which is the AH state. The nearest-neighbor interaction, for example, favors the chiral symmetry breaking, and is believed to be relevant in the true magnetic field \cite{goerbig}; in the case of axial field, however, the chiral symmetry is broken already by the field, and the only channel available to the system to lower its energy  is the breaking of the remaining time reversal symmetry, i. e. the AH state.

The interacting Hamiltonian reads
\begin{equation}
H=\sum_{\langle \alpha, \beta\rangle } t_{\alpha \beta} \left( c^\dagger_\alpha c_\beta + c^\dagger_\beta c_\alpha \right)+V \sum_{\langle \langle \alpha,\beta \rangle \rangle} n_\alpha n_\beta .
\label{interaction}
\end{equation}
$\langle\langle \dots \rangle\rangle$ stands for the sum over the nearest-neighbor and second-neighbor sites, and $n_\alpha$ is the fermion occupation number on site $\alpha$. The usual Fock decomposition yields an effective single-particle Hamiltonian
\begin{equation}\label{SPHamil}
H_{SP}=\sum_{\langle \alpha, \beta \rangle} t_{\alpha \beta} \left( c^\dagger_\alpha c_\beta + c^\dagger_\beta c_\alpha \right) + \sum_{\langle \langle \alpha, \beta \rangle\rangle} ( \eta_{\alpha \beta } c^{\dagger}_\beta c_\alpha + \mathrm{H.c} ).
\end{equation}
 We will assume the intrasublattice circulating current $\eta_{\alpha\beta}=V \langle c^{\dagger}_\alpha c_\beta \rangle$ to be purely \emph{imaginary}, and oriented in opposite directions on the two sublattices. It will play the role of the TRS breaking AH order parameter\cite{haldane}, and we proceed to determine it self-consistently on a finite honeycomb lattice. In the presence of axial fields only the AH gap can split the zero-energy subspace and the OP $\langle \Psi^\dagger i \gamma_1 \gamma_2 \Psi \rangle$ transforms as a \emph{scalar} under the chiral rotation.
\begin{figure}[htb]
\hspace{-0.2cm}\includegraphics[width=6.5cm,height=4.25cm]{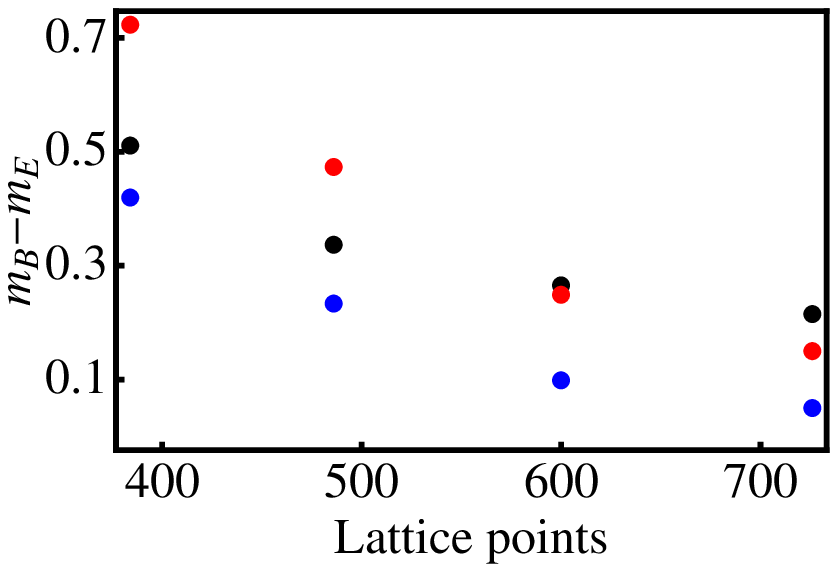}\\
\includegraphics[width=7.0cm, height=4.75cm]{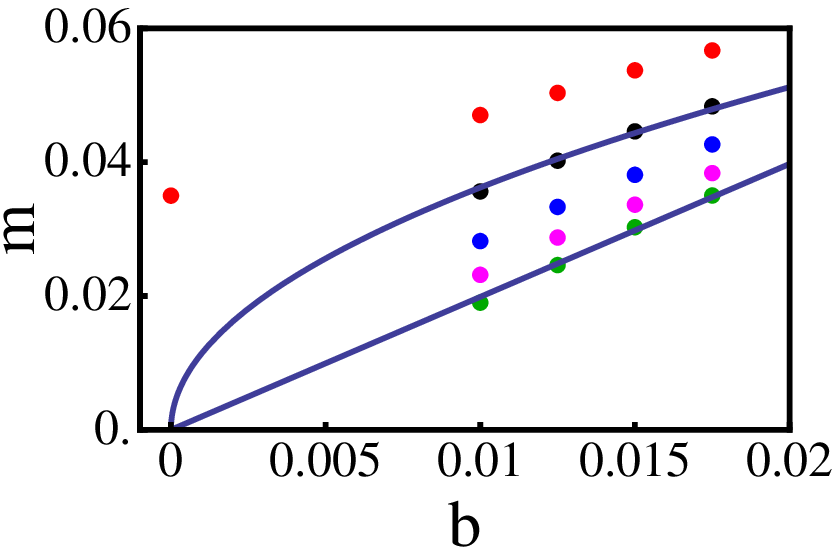}
\caption{(Color online) Top: Difference of the AH OP (in units of $t$) in bulk ($m_B$) and edge ($m_E$) on A-sublattice as a function of the system size for V=1.27(red), 1.0(black), 0.75(blue) with $b=0.015 b_0$. Bottom: Scaling of anomalous Hall OP with axial magnetic fields ($b/b_0$). The red, black, blue, magenta and green dots respectively correspond to $V=1.5, 1.27 (V_C),1.0,0.75.0.5$.}
\label{scaling}
\end{figure}
In the absence of an axial field ($\chi\equiv 0$) a non-zero self-consistent solution of the AH order $\eta$ on a honeycomb lattice of 600 sites is  found only for $V > 1.27$ \cite{Supplementary}. The amplitudes of the OP on both sublattices are then equal. For $V<V_c$, the self-consistent value of $\eta$ vanishes everywhere in the system. Hence, we can define $V_c=1.27$ as the zero-field critical interaction. This value is very close to the one found analytically. \cite{raghu, franz}

After the introduction of the uniform axial field in the same system, we  search for the self-consistent solution of $\eta$ at sub-critical interactions $V < V_c$ as well. Typical distributions of the OPs then are shown in Fig.~\ref{AHlattice}. For $V<V_c$, the TRS breaking OP on the sublattice A (B) clearly forms in the bulk (boundary) of the system. This spatial separation of the OPs on two sublattices follows the structure of the near-zero-energy states. The OP on the sublattice A is roughly uniform in the bulk (for $ r \leq 6$), and with the inhomogeneity disappearing with an increase of the  system size (see Fig.~\ref{scaling}, top). As the interaction gets stronger ($V \sim V_c$), the effect of the axial field becomes irrelevant and OPs on two sub-lattices become comparable; see the red, black, and blue curves in Fig. \ref{AHlattice}. The effective single-particle Hamiltonian $H_{SP}$ in Eq. (~\ref{SPHamil}) preserves the $C_3$ symmetry of the bond configuration in Fig. 1.

\begin{figure}[htb]
\includegraphics[width=7.95cm,height=5.26cm]{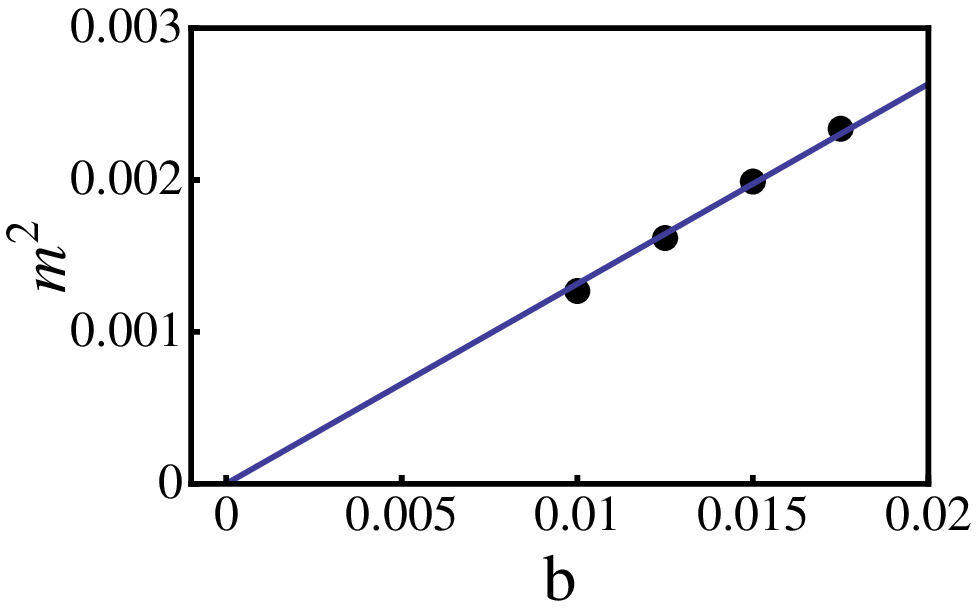}\vspace{0.4cm}\\
\includegraphics[width=7.5cm,height=5.5cm]{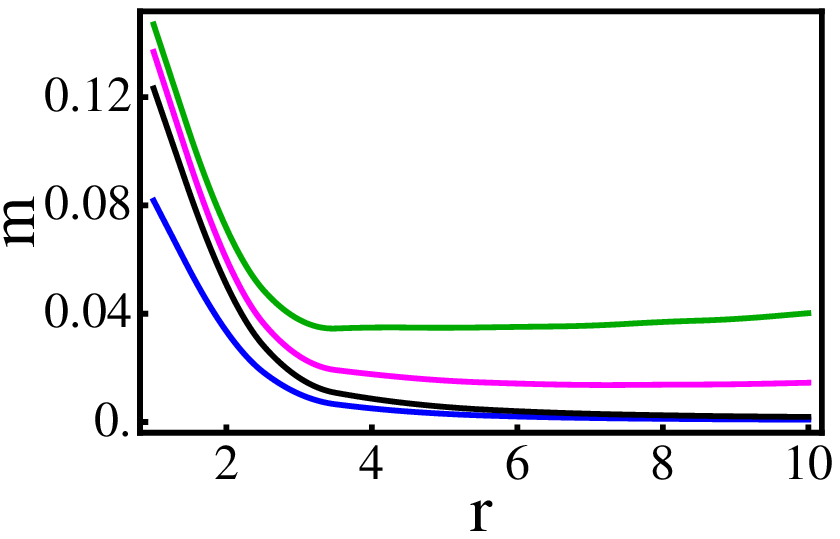}
\caption{(Color online) Top: Linear scaling of $m^2$ with $b$ for $V=V_c= 1.27$. Bottom: TRS breaking OP for V=1.5 (green), 1.27 (magenta), 1.0 (black), 0.75 (blue) with total flux $7.85 \Phi_0$ of a bell-shaped nonuniform axial field. $\Phi_0=h/e$ is flux quantum.}
\label{inhomog}
\end{figure}

The OP on the sublattice A (averaged up to $r=6$) scales linearly with the uniform axial field (b) when $V \ll V_c$ (Fig. \ref{scaling}, bottom). The scaling becomes sub-linear for intermediate strength of the interaction. At the zero-field criticality ($V=1.27$), the gap appears to scale as $\sqrt{b}$. This behavior can also be confirmed from the linear scaling of $m^2$ with $b$ which passes though the origin, as shown in Fig. \ref{inhomog} (top). Finally, for $V > V_c$ the OP saturates to a finite value at zero field. With our definition of the first LL energy $E_1$, the \emph{universal} ratio of $E_1$ to the mass gap at $V=V_C$ is found here to be $\approx 5.78$. The scaling of the TRS-breaking OP is thus similar to the scaling of the chiral-symmetry-breaking OP with the real magnetic field found previously \cite{bitan-numerics, bitan-scaling}. The universal ratio is also reasonably close to the value $5.985$, obtained analytically for the chiral-symmetry-breaking OP.\cite{bitan-scaling}

Existence of the near zero energy states even when the axial magnetic field assumes a spatially non-uniform profile, allows the formation of the AH order for $V \ll V_c$. With a \emph{bell-shaped} field, localized around the center of the system, a typical distribution of the OP on the sublattice A, computed self-consistently on a $726$ site honeycomb lattice, is shown in the bottom of Fig.~\ref{inhomog}. For $V \ll V_c$, the OP forms only in the vicinity of the localized flux. This behavior of the OP on the A sub-lattice is generic and the OP disappears towards the boundary of larger systems\cite{Supplementary}. However, for sub-critical interactions the OP is also accompanied by a \emph{real amplitude}, which appears to be a finite size effect\cite{Supplementary}. On the other hand, when $V \sim V_c$, the OP starts to become considerable everywhere in the system. Behavior of the OP on the sublattice B is qualitatively similar to that in the presence of uniform axial field.

\section{Competition between anomalous Hall and anomalous spin Hall insulators}

Let us now restore the spin degrees of freedom, and define the eight-component spinor $\Psi=(\Psi_\uparrow,\Psi_\downarrow)$, where $\sigma=\uparrow, \downarrow$ are the spin projections along the $z$-axis. The four-component spinors $\Psi_\sigma$ for each spin projection take the form of $\Psi$ in Eq. (\ref{spinor}). The Dirac Hamiltonian in this basis is $H_D[a]=\sigma_0 \otimes H[a]$. The finite range components of the Coulomb interaction then allow the formation of various other ordered phases besides the AH insulator; for example, the anomalous SH insulator, or the spin-polarized ferromagnetic state. For example, the on-site Hubbard repulsion $U$ typically favors the spin-polarized state \cite{ashwin}. On the other hand, the second-nearest- neighbor repulsion prefers the AH or the SH insulators \cite{herbut-pseudocatalysis, raghu, abanin}. Depending on the relative strengths of the finite ranged components of the Coulomb interaction there will be a competition among various ordered states in strained graphene. In this work, however, we will take only the second-nearest-neighbor repulsion into account, and study the competition between the AH and SH insulators. Possible appearance of the ferromagnetic ground state for  strong on-site Hubbard $U$, and its competition with the AH and the SH orders will be addressed in a separate publication.

The SH and the AH OPs are $\vec{C}=\langle \Psi^\dagger \; \left[ \vec{\sigma} \otimes i \gamma_1 \gamma_2 \right] \Psi \rangle$, and $C_0=\langle \Psi^\dagger \; \left[ \sigma_0 \otimes i \gamma_1 \gamma_2 \right] \Psi \rangle$, respectively. The SH insulator is even under the TRS, which now also includes the usual reversal of the spin, whereas the AH state is odd. \cite{totalTRS} Notice that the matrices appearing in both the AH and SH OPs anticommute with the Dirac Hamiltonian  $H_D[a]$. Hence, both of these OPs upon developing a finite expectation value would not only split the states at zero energy, but would also shift downward the occupied states with negative energies. For example, if the axial field is uniform, the LLs at $\pm \sqrt{2 n b}$ get pushed to $\pm \sqrt{2 n b + X^2}$, where $X=C_0$ or $|\vec{C}|$. Hence, at half filling when all the states at negative (positive) energies are filled (empty), it is energetically highly  advantageous to develop such a ``mass gap."
\begin{figure}[htb]
\includegraphics[width=4.9cm,height=5.0cm]{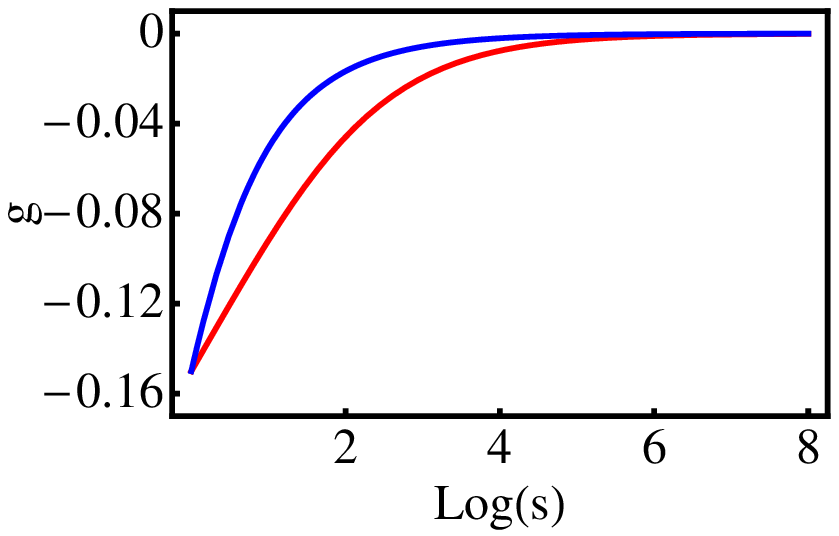} \hspace{0.05cm}
\includegraphics[width=3.50cm,height=5.0cm]{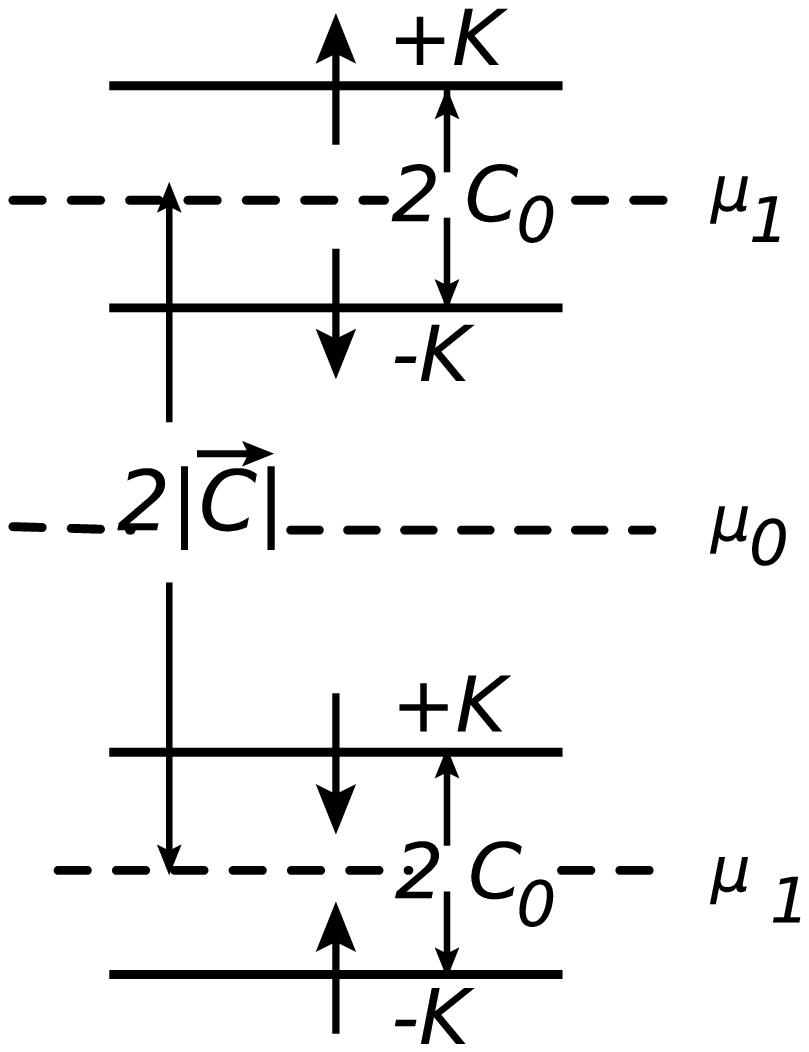}
\caption{(Color online) Left: Flow of the couplings $g_c$(red), and $g_h$(blue). Right: Proposed splitting of the zero energy subspace with both SH and AH order present. All the states are localized on one sub-lattice. Both SH and AH order parameters are formed when the chemical potential is at $\mu_1$.}
\label{AHQSHfig}
\end{figure}

Let us now address the competition between the AH and the SH insulators in the following way.
Consider first the interaction Lagrangian in the continuum, and at zero field, as
\begin{equation}
L_{int}=g_h \left( \Psi^\dagger \sigma_0 \otimes i \gamma_1 \gamma_2 \Psi\right)^2 + g_c \left( \Psi^\dagger \vec{\sigma} \otimes i \gamma_1 \gamma_2  \Psi\right)^2.
\end{equation}
The interactions $g_h < 0$ and $g_c < 0$ favor the AH and SH insulators, respectively. In graphene at $b=0$ any weak electron-electron interaction is irrelevant,  and to place the system in an ordered phase the interaction needs to be sufficiently strong\cite{herbut-interaction, kekule}. Assuming only the second-nearest-neighbor repulsion $V>0$, $g_h, g_c \sim - V$ at the lattice scale, and they are equally irrelevant if $V/t \ll 1$. At the mean-field level therefore the AH and the SH states are degenerate. As one integrates out the fast Fourier modes in momentum shell $\Lambda/s < |k| <\Lambda$, where $s > 1$; however, these two couplings flow to zero differently. This is a consequence of the fact that the corresponding OPs obviously break different symmetries, one of which is discrete, and the other continuous. To the second order, the flow of these two couplings is given by the $\beta$-functions ($\beta_{X}=d X / d \log{s}$)
\begin{equation}
\beta_{g_h}=-g_h(1 + 3 g_h - 3 g_c), \beta_{g_c}=-g_c(1 +5 g_c - g_h),
\end{equation}
after rescaling $(2\Lambda/\pi^2) g_{h,c} \rightarrow g_{h,c}$. \cite{newcoupling}

 Negative linear terms in the $\beta$-functions imply that any spontaneous symmetry breaking in graphene at zero field can occur only at strong interactions. In the presence of an axial or standard magnetic field, however, the ordering will occur even at weak interactions, due to the zero-energy states, as we have argued above. When the axial field is roughly uniform we will choose the final value of the parameter $s \sim l_b /a \gg 1$, where $a$ is lattice spacing and $l_b \sim 1/\sqrt{b}$, and determine the effective values of the couplings at the  scale of the \emph{axial magnetic length}. We therefore numerically solve the coupled flow equations with $g_c=g_h=V$ when $s=1$. The less irrelevant coupling at the scale $s \sim l_b /a \gg 1$ will then give rise to the dominant instability in strained graphene. The flows of the coupling constants $g_h$ and $g_c$ are shown in Fig.~\ref{AHQSHfig} (left), and one can see that $g_c$ is less irrelevant than $g_h$ for any $s>1$. The leading instability at weak coupling in the presence of the axial field is therefore the SH state. The same outcome is also found at strong coupling  and at zero field, in agreement with the previous study. \cite{raghu}
 
 When the chemical potential is close to the first excited states at $\pm |\vec{C}|$ the system can develop a gap by breaking the TRS and by  developing an additional AH order, as shown in Fig.~\ref{AHQSHfig} (right). The resulting single-particle excitation gap of the AH state is then $2 C_0$, and the zero-energy LL is 1/4 or 3/4 full. The scaling of the AH OP, which one can possibly realize in strained graphene  in this way and away from the neutrality point is qualitatively similar to the one we have computed numerically at the neutrality point. When the axial field is nonuniform the same mechanism still operates, with the AH OP then developing mostly in the vicinity of the localized flux at weak interaction.

\section{Summary and discussion}

To summarize, we proposed a specific modulation of the nearest-neighbor hopping amplitude that captures the effect of time reversal symmetric axial magnetic fields. Although such a field can also be spatially nonuniform, it always produces a finite number of states near zero energy. We show that such a spectrum is conducive to formation of topologically non-trivial ordered phases such as the AH and SH insulators, even for weak repulsive interactions. We provide numerical evidence that spinless fermions can be in the AH phase even at weak second-nearest-neighbor repulsive interaction, with the magnitude of the OP depending linearly on the axial field. In the competition between the AH and SH states at weak interaction and in finite uniform axial field we found that at half filling the SH state wins. The TRS breaking AH order parameter, however, appears at finite doping. We hope that the realization of time reversal symmetric axial field and its tunability over a wide range in real \cite{castroneteoRu} and artificial graphene \cite{moleculargraphene} will make these effects soon visible in experiments.

\section{Acknowledgement}

B. R. was supported at National High Magnetic Field Laboratory by NSF cooperative agreement No. DMR-0654118, the State of Florida, and the U. S. Department of Energy. I. F. H. was supported by the NSERC of Canada.

\pagebreak

\begin{widetext}

\begin{center}
{\bf \large Supplementary materials for ``Topological insulators in strained graphene at weak interaction"}\\
{Bitan Roy$^{1}$, Igor F. Herbut$^{2,3}$}\\
{$^1$ \emph{National High Magnetic Field Laboratory and Department of Physics, Florida State University, FL 32306, USA}} \\
{ $^2$ \emph{Max-Planck-Institut f\"ur Physik Komplexer Systeme, N\"othnitzer Str. 38, 01187 Dresden, Germany}} \\
{ $^3$ \emph{Department of Physics, Simon Fraser University, Burnaby, British Columbia, Canada V5A 1S6}}
\end{center}

\end{widetext}

\vspace{10pt}

In the paper we have presented numerical evidence (for spinless fermions) that in the presence of axial magnetic fields the TRS can be spontaneously broken at weak second-nearest-neighbor repulsion, so that the system enters into the AH insulator phase. Below we provide some additional numerical results in support of our claim.
\begin{figure}[htb]
\begin{center}
\includegraphics[width=7cm,height=4.cm]{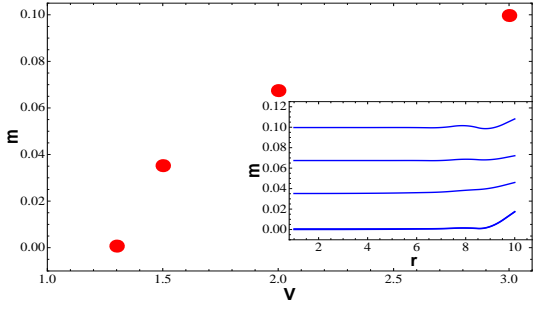}
\end{center}
\caption{Variation of the TRS-breaking AH order with the next-nearest-neighbor interaction ($V$) at zero axial field. Inset: The variation of AH order in the entire system on $A$ and $B$ sub-lattices for $V=3,2,1.5,1.3$, reads from top to bottom.
 }\label{Fig-critint}
\end{figure}

\subsection{Zero field criticality}

Let us first present the numerical solution of AH OP for zero axial field and estimate the critical strength of interaction ($V_c$) for insulation. In the paper we have mentioned that in the absence of axial fields, the next-nearest-neighbor component of the Coulomb interaction ($V$) in graphene needs to be sufficiently large to develop a TRS-broken state. A nonzero, and fairly uniform self consistent solution of the AH OP, computed on a $600$ site honeycomb lattice can only be found for $V >1.27$, as shown Fig. \ref{Fig-critint}. The magnitude of the OP on both sublattices is equal. For $V<1.27$ the OP vanishes everywhere in the system, which is an artifact of the self-consistent Fock approximation. The value of $V_c$ does not depend on system's size, when it contains more than $600$ lattice points. We can then define $V=1.27$ as the critical strength of interaction for the insulation in the absence of axial fields.
\begin{figure}[htb]
\begin{center}
\includegraphics[width=3.5cm,height=2.5cm]{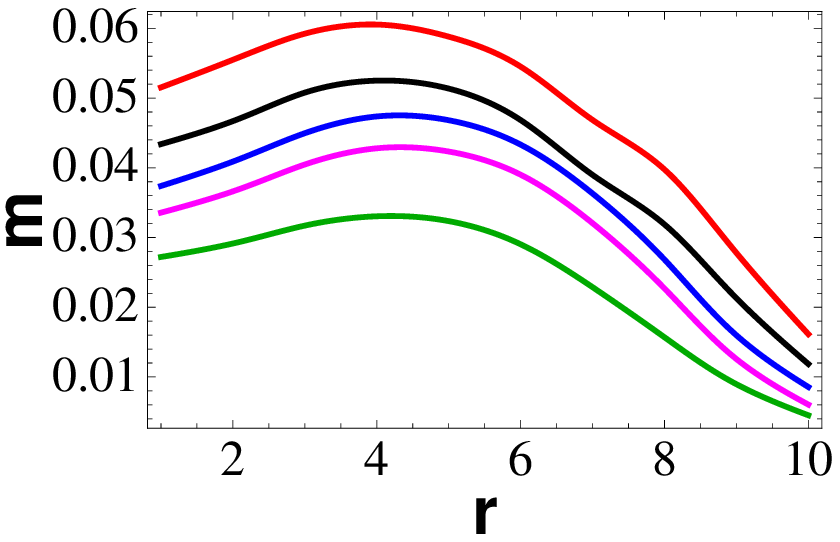} \hspace*{0.25cm}
\includegraphics[width=3.5cm,height=2.5cm]{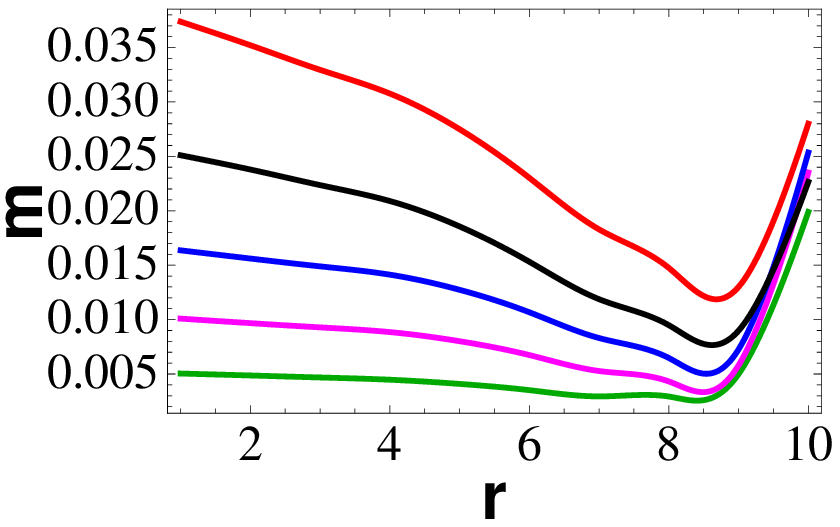}\\ \vspace*{0.2cm}
\includegraphics[width=3.5cm,height=2.5cm]{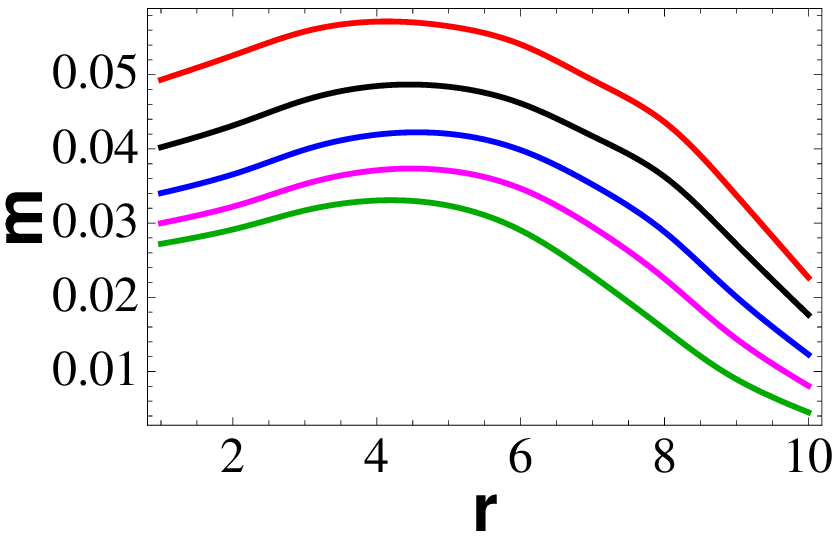}\hspace*{0.25cm}
\includegraphics[width=3.5cm,height=2.5cm]{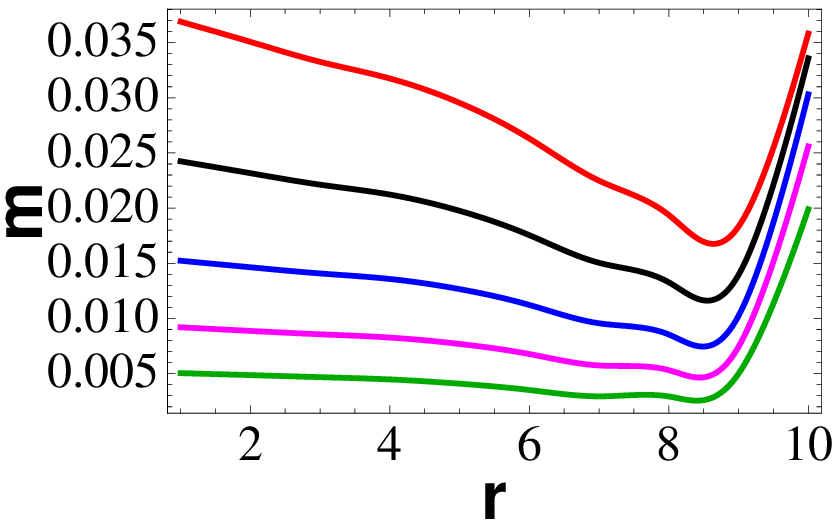}
\end{center}
\caption{Self-consistent solution of the AH order on A (left) and B (right) sublattices of a honeycomb lattice of 600 sites, in the presence of uniform axial magnetic field with $b=0.035 b_0$ (top), $0.03 b_0$ (bottom). The red, black, blue, magenta, green curves correspond to $V=1.5,1.27,1.0,0.75,0.5$, respectively.
 }\label{suppleuniform}
\end{figure}

\subsection{Pseudo magnetic catalysis}
In the paper, we have presented the distribution of the AH OP on A and B sub-lattices for a particular strength of the uniform axial magnetic field. However, we have performed the numerical analysis for various other strengths of the uniform axial magnetic field. In Fig. \ref{suppleuniform}, we present the AH OP on two sub-lattices for two different strengths of uniform axial magnetic field.
\begin{figure}[htb]
\begin{center}
\includegraphics[width=3.95cm,height=3.05cm]{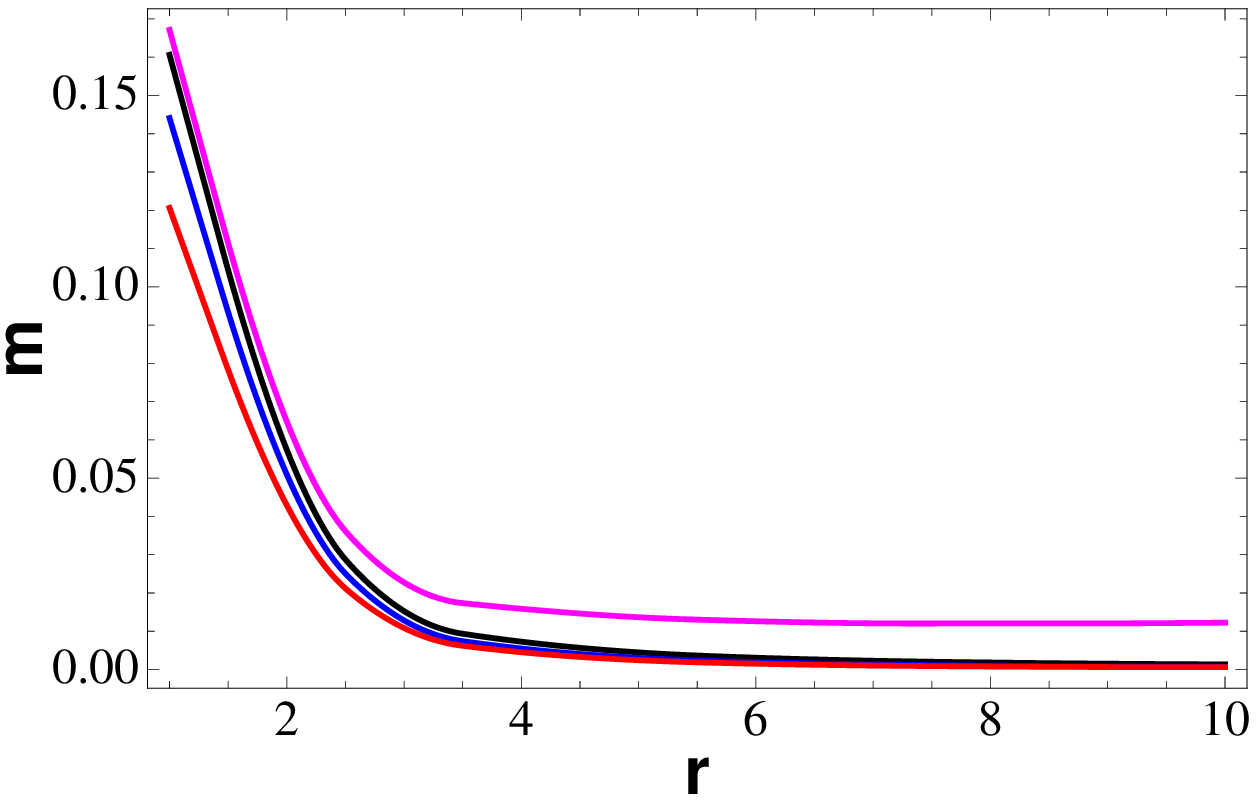}\hspace*{0.25cm}
\includegraphics[width=3.95cm,height=3.05cm]{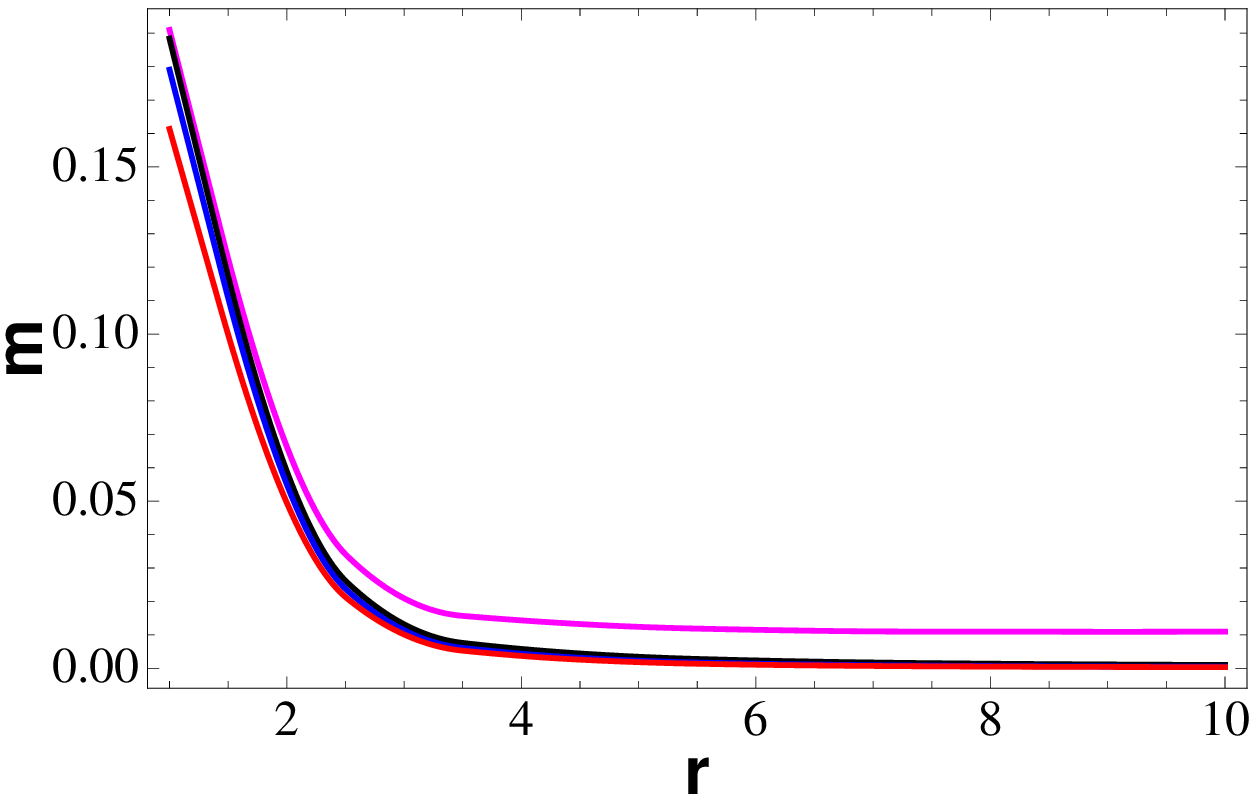}
\end{center}
\caption{Self consistent AH OP on the honeycomb lattice of $726$ sites, in the presence of a nonuniform magnetic field with total axial flux $\Phi_{total}=9.42 \Phi_0$ (left), $10.99 \Phi_0$ (right). The strength of the interaction reads as $V=1.27, 1, 0.75, 0.5$ from top to bottom.
 }\label{supplenonuniform}
\end{figure}

We have also searched for the self-consistent solution of AH order in the presence of a bell-shaped axial magnetic field, localized around the center of the system. In the paper we have presented the self-consistent solution of the AH OP on the sublattice A, for a particular choice of the total axial flux passing through the system. However, the same analysis has been performed for other values of the total axial flux penetrating the system, and the results are shown in Fig. \ref{supplenonuniform}.
\begin{figure}[htb]
\begin{center}
\includegraphics[width=6cm,height=4.cm]{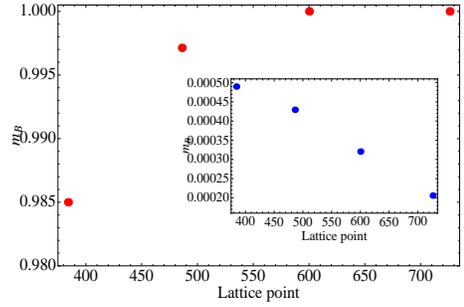}
\end{center}
\caption{AH OP in the vicinity of the localized flux as a function of the system size for $V=1.0$, and total flux $7.85 \Phi_0$. Here we have normalized the OP with respect to the maximum one, obtained on a $726$ site honeycomb lattice. Inset: variation of the OP far away from the localized flux. }
\label{OPsatures}
\end{figure}

\subsection{Finite size effects in non-uniform condensation}
\begin{figure}[htb]
\begin{center}
\includegraphics[width=6cm,height=4.cm]{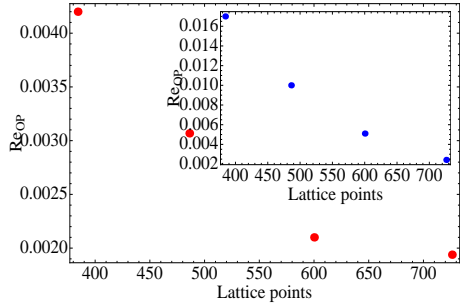}
\end{center}
\caption{ Variation of the the real component of the AH OP in the vicinity of the localized flux with the system size, for $V=1.0$, and when a total flux $7.85 \Phi_0$. Inset: the same quantity far from the localized flux.
 }\label{Resys}
\end{figure}

In the paper, we have shown that in the presence of a bell-shaped axial magnetic field, localized around the center of the system, the AH OP  dominantly develops in the vicinity of the localized axial flux, when the interaction is sub-critical. Far away from the localized field, the OP disappears. Fig. \ref{OPsatures} shows that such behavior is generic, and the OP very much saturates in the bulk of the honeycomb lattice of $726$ sites, whereas that near the boundary of the system gradually disappears as the system size is increased. Furthermore, we have noticed that the AH OP in the presence of a non-uniform axial magnetic field is accompanied by a real component, for sub-critical interactions. In Fig. \ref{Resys} shows that appearance of such real component appears to be a finite size effect.


\begin{thebibliography}{99}
\bibitem{totalTRS} TRS operator for spinful fermions in our representation
is $\sigma_2 \otimes i \gamma_1 \gamma_5 \; K$.
\bibitem{newcoupling} In the one-loop renormalization group calculation we also generate new quartic terms, $ g(\Psi^\dagger \vec{\sigma} \otimes \gamma_0 \gamma_\mu \Psi)^2$, where $\mu=0,1,2$. They are, however,  more irrelevant.
\end{thebibliography}
\end{document}